\documentclass[final,3p,times,twocolumn]{elsarticle}

\usepackage{amssymb}
\usepackage{amsmath}
\usepackage{enumitem}
\setlist{nolistsep}
\usepackage{upquote}
\usepackage{hyperref}
\usepackage[dvipsnames]{xcolor}
\usepackage{listings}
\newcounter{bla}

\journal{Computer Physics Communications}

\begin{document}

\begin{frontmatter}



\title{NGAMMA: A Monte Carlo generator for multiphoton \\production in $e^+e^-$ annihilation}


\author[a,b]{L.~V.~Kardapoltsev}
\author[a]{N.~A.~Melnikova}

\cortext[author] {Corresponding author.\\\textit{E-mail address:} l.v.kardapoltsev@inp.nsk.su}
\address[a]{Budker Institute of Nuclear Physics, SB RAS, Novosibirsk, 630090, Russia}
\address[b]{Novosibirsk State University, Novosibirsk, 630090, Russia}

\begin{abstract}
We present the NGAMMA Monte Carlo event generator for QED processes of $e^+e^-$ annihilation into 
a multiphoton final state, $e^+e^-\to N\gamma (N \ge 2)$.
These processes are an important source 
of background in the study of $e^+e^-\to$\textit{hadrons} processes with a multiphoton final state,
especially for experiments at low energy $e^+e^-$ colliders like SND, CMD-3, KLOE 
and for the BESIII experiment.
For generation, NGAMMA exploits the exact tree-level amplitude.



\noindent \textbf{PROGRAM SUMMARY}

\begin{small}
\noindent
{\em Program Title:} NGAMMA                                          \\
{\em CPC Library link to program files:} (to be added by Technical Editor) \\
{\em Developer's repository link:} \url{https://git.inp.nsk.su/ngamma/ngamma} \\
{\em Licensing provisions(please choose one):} LGPL\\ 
{\em Programming language:} C++                                   \\
{\em Nature of problem:} The NGAMMA generator was developed to simulate the background for
$e^+e^-\to$\textit{hadrons} processes with a multiphoton final state
coming from the QED processes $e^+e^-\to N\gamma (N \ge 2)$.\\
{\em Solution method:} Events consisting of momenta of the outgoing particles are generated
by Monte Carlo methods. The generated events are distributed according to
the exact tree-level cross section~\cite{1}.



\end{small}    
\end{abstract} 
\end{frontmatter}

\section{Introduction}
\label{intro}
Processes of quantum electrodynamics (QED) such as 
multiphoton annihilation of an $e^+e^-$ pair
\begin{equation}
e^+e^-\to N\gamma~(N \ge 2),
\label{Ngamma}
\end{equation}
are an important source of background in the study of the
$e^+e^-\to$\textit{hadrons} processes with a multiphoton final state. 
For example, taking them into account is important when studying processes
with a five-photon final state
$e^+e^-\to \pi^0\pi^0\gamma$, $e^+e^-\to \eta\pi^0\gamma$ with the decay of $\pi^0$ and $\eta$ into two photons.

The total cross section of the process~(\ref{Ngamma}), obtained in the double-logarithmic approximation,
looks as follows~\cite{Gorshkov_Lipatov}
\begin{equation}
\sigma_{N} = \frac{2\alpha^2\pi\rho}{s}\left(\frac{\alpha\rho^2}{\pi}\right)^{N-2}\frac{2^{3-N}}{N!(N-2)!}, 
\label{NgammaCrsLL}
\end{equation}
where $s = (p_{+}+p_{-})^2 = E^2$, $E$ is the center of mass energy of the initial 
$e^+e^-$ pair, $\rho=ln(s/m_e^2)$, $m_e$ is the electron mass,
$\alpha = e^2/4\pi$ is the fine structure constant. From this expression, 
one can see that the cross section decreases rapidly with the increasing
center of mass energy $E$ and the number of photons in the final state $N$.
For this reason, taking into account the processes~(\ref{Ngamma}) 
is more important for $e^+e^-$ colliders of relatively low energies, such as
VEPP-2000~\cite{vepp2k} and $\rm DA\Phi NE$~\cite{DAFNE}. 
They also should be considered
when studying rare processes with a multiphoton final state at BESIII~\cite{BESIII}.
For illustration, the cross sections of the processes~(\ref{Ngamma})
are shown in Tab.~\ref{CrsTab}. They are calculated with
restrictions on the energies $E_{\gamma}$ and polar 
angles $\theta_{\gamma}$ of photons, corresponding to their detectability
in the SND detector~\cite{SND_desc1, SND_desc2, SND_desc3, SND_desc4} collecting data at the VEPP-2000 collider

\begin{equation}
E_{\gamma} > 20~\rm{ MeV}, \qquad    18^{\circ} < \theta_{\gamma} < 162^{\circ}
\label{SNDcut}
\end{equation}
and in the BESIII detector
\begin{equation*}
E_{\gamma} > 25~\rm{MeV}~\rm{ for }~|{\rm cos}\theta_{\gamma}| < 0.8 \qquad~\rm{ and }
\end{equation*}
\begin{equation}
E_{\gamma} > 50~\rm{ MeV }~\rm{ for }~0.86 < |{\rm cos}\theta_{\gamma}| < 0.92.
\label{BEScut}
\end{equation}

\begin{table*}[t!]
\centering
\begin{tabular}{|c|c|c|c||c|c|c|c|}
\hline
$E$, GeV & $\sigma_{4\gamma}^{\rm SND}$, pb & $\sigma_{5\gamma}^{\rm SND}$, pb & $\sigma_{6\gamma}^{\rm SND}$, fb & $E$, GeV & $\sigma_{3\gamma}^{\rm BESIII}$, nb & $\sigma_{4\gamma}^{\rm BESIII}$, pb & $\sigma_{5\gamma}^{\rm BESIII}$, pb  \\  \hline
0.782 & 470 & 5.8 & 50 & 2.0   & 2.2 & 41 & 0.48  \\  
1.020 & 343 & 4.8 & 47 & 3.096 & 1.1 & 22 & 0.31  \\ 
2.0   & 141 & 2.6 & 33 & 3.773 & 0.76 & 17 & 0.25  \\ 
\hline
\end{tabular}
\caption{\normalsize 
Total cross sections for the processes $e^+e^- \to 3\gamma, 4\gamma, 5\gamma, 6\gamma$
calculated with restrictions on parameters of the final photons corresponding to their detectability in the SND and BESIII detectors.
\label{CrsTab}}
\end{table*}

The reaction $e^+e^-\to \gamma\gamma$ is widely used for luminosity measurement.
For this reason, there are many well-established  Monte Carlo generators for this reaction:
BabaYaga@NLO~\cite{BabaYaga}, MCGPJ~\cite{MCGPJ1, MCGPJ2} and BKQED~\cite{BKQED}. 
All of them are well suited for generating three photon production as well.
Although the BabaYaga@NLO generates an arbitrary number of photons in the final state, four and more photons 
are generated according to the cross section in the leading logarithmic approximation. 
It provides insufficient accuracy for studying the background.
To generate processes~(\ref{Ngamma}) with the exact Born amplitude, one can use,
for example, the CompHEP package~\cite{CompHEP}. It allows to calculate 
the matrix element of a QED process symbolically and then generate events
according to it. Unfortunately, CompHEP has a very low event generation performance,
especially for five and more photons in the final state.
The main goal of developing the NGAMMA generator is to provide a Monte Carlo generator for
the processes~(\ref{Ngamma}) with $N \ge 4$, which uses the exact Born amplitude
and has a sufficiently high generation performance.

\section{Born amplitude}

The differential cross section of
\begin{equation}
e^+(p_+)+e^-(p_-)\to\gamma(k_1)+\gamma(k_2)+{\ldots} +\gamma(k_N)
\end{equation}
can be expressed as follows
\begin{equation}
d\sigma_{N} = \frac{1}{F}\cdot\left(d\Phi_{N}\right)\cdot\frac{1}{N!}\cdot\frac{1}{4}\cdot\sum_{S}\left|\sum_{D}M(D,S)\right|^2,
\label{NgammaCrs1}
\end{equation}
where $F = 2E^2$ is the flux factor, $d\Phi_{N}$ is the phase space volume, $\sum_{S}$ is the sum over photon polarization states 
$\lambda_i = +$ or $-$, $\sum_{D}$ is the sum over the final photon permutations. 
In the limit of massless fermions, the amplitudes corresponding to opposite helicities of the electron and positron are equal to zero.
The amplitudes corresponding to the same helicities coincide, therefore we need to consider only one of them.
The exact expression for this amplitude in the Born approximation is obtained in Ref.~\cite{Kleiss_ng} 
\begin{equation*}
M = e^N \left( \prod_{j=1}^{N} p\cdot k_{j} \right)^{-1/2} s(p_{+},a_1)t(b_N,p_{-})
\end{equation*}
\begin{equation}
\times \prod_{i=1}^{N-1}\left[\frac{\zeta_i t(b_i,\hat{q}_i)s(\hat{q}_i,a_{i+1})}{q_i^2} + \frac{ t(b_i,p)s(p,a_{i+1})}{2 p\cdot q_{i}} \right],
\label{NgammaM}
\end{equation}
where
\begin{equation*}
a_i = p,~b_i = k_i~{\rm if } ~\lambda_i = +~ {\rm and } ~a_i = k_i,~b_i = p~{\rm if }~ \lambda_i = -,
\end{equation*}
\begin{equation*}
s(p_1,p_2) = (p_1^y+ip_1^z)\sqrt{\frac{p_2^0-p_2^x}{p_1^0-p_1^x}} - (p_2^y+ip_2^z)\sqrt{\frac{p_1^0-p_1^x}{p_2^0-p_2^x}},
\end{equation*}
\begin{equation*}
t(p_1,p_2) = [s(p_2,p_1)]^* \qquad q_i = \sum_{j=1}^{i}k_j - p_{+},
\end{equation*}
\begin{equation*}
\hat{q}^{\mu}_i=\zeta_i\left(q^{\mu}_i-\frac{q^2}{2 p\cdot q_i}p^{\mu}\right), \qquad \zeta_i = sign\left(q_i^0-\frac{q^2}{2 p\cdot q_i}p^{0}\right),
\end{equation*}
$p$ is an arbitrary light-like four-vector not collinear with $k_i$. The result is independent of the choice of the four-vector $p$, so we 
take it as equal to $p = p_{+}$ because in this case half of the amplitudes $M(D,S)$ vanish in the sum over $S$ in equation~(\ref{NgammaCrs1}).

\begin{figure}[h!b]
\center
\includegraphics[width=1.1\columnwidth]{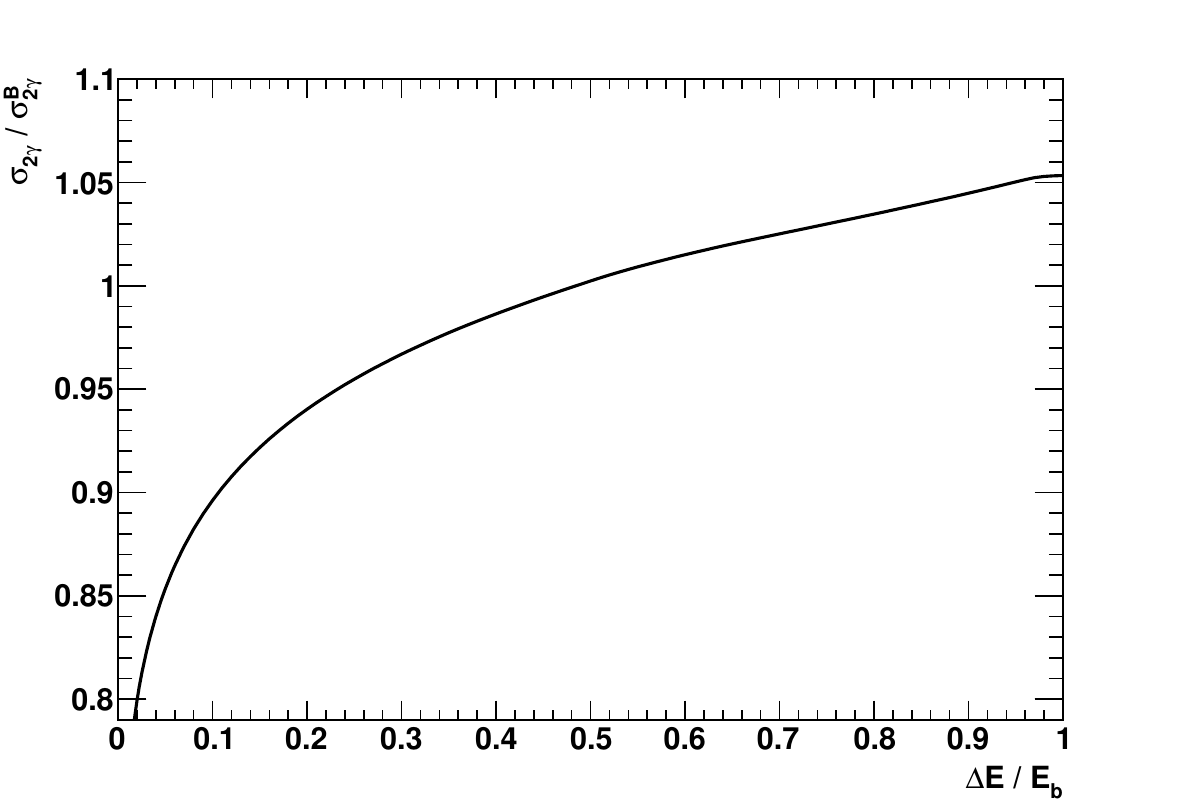}
\caption{\label{Crs2g}
The ratio of the $e^+e^- \to 2\gamma$ cross sections calculated using BabaYaga@NLO ($\sigma_{2\gamma}$)
and NGAMMA ($\sigma_{2\gamma}^{B}$).
The $\sigma_{2\gamma}$ is calculated with a 
constraint on the total energy of the additional photons in the event to be less than $\Delta E$.
See text for details.
}
\end{figure}

The NGAMMA generator exploits the expressions~(\ref{NgammaM}) obtained in the Born approximation
to describe the $e^+e^- \to N\gamma$ cross section.
To evaluate the theoretical accuracy of this approach,
we made a comparison with the BabaYaga@NLO generator~\cite{BabaYaga}.
For the process $e^+e^- \to 2\gamma$, this generator takes into account next-to-leading order QED corrections and the emission of an arbitrary number of photons
in the leading logarithmic approximation. Therefore, for $N = 2$, it provides much better precision than the Born approximation. 
For $N = 3$, it uses the Born cross section and provides comparable precision. In this case, the comparison is useful for 
evaluating the influence of additional radiation.

Using BabaYaga@NLO, we generated a sample of $e^+e^-\to 2\gamma$ events at the c.m. energy $E = 2$ GeV.
Since the accuracy of the Born approximation is highly dependent on the restrictions 
imposed on the kinematics of the event, we need to emulate those restrictions as well.
For this, we selected only those generated events for which at least two photons 
satisfy conditions~(\ref{SNDcut}) and the total energy of all photons in an event, 
except for the two most energetic ones satisfying conditions~(\ref{SNDcut}), is less than $\Delta E$.
The $e^+e^- \to 2\gamma$ cross section, integrated over 
the phase space defined by these cuts, was calculated as follows

\begin{equation}
\sigma_{2\gamma} = \sigma^{0}_{2\gamma}\cdot N_{sel} / N_{0},
\end{equation}

where $\sigma^{0}_{2\gamma}$ is the cross section calculated by BabaYaga@NLO for the whole generated sample, 
$N_{sel}$ is the number of events passed the selection conditions, and
$N_{0}$ is the total number of generated events. Figure~\ref{Crs2g} shows the dependence 
of $\sigma_{2\gamma} / \sigma_{2\gamma}^{B}$ on $\Delta E / E_{b}$, where  $E_{b}=E/2$ and
$\sigma_{2\gamma}^{B}$ is the Born cross section integrated over the phase space defined by conditions~(\ref{SNDcut}). 
For most experimentally accessible kinematic restrictions $\Delta E / E_{b} \gtrsim 0.1$,
the difference between $\sigma_{2\gamma}$ and $\sigma_{2\gamma}^{B}$ is $\lesssim 10 \%$.
But for very tight restrictions on the kinematics $\Delta E / E_{b} \approx 0$, 
the Born approximation~(\ref{NgammaM}) is not applicable.

\begin{figure}[ht!]
\center
\includegraphics[width=1.1\columnwidth]{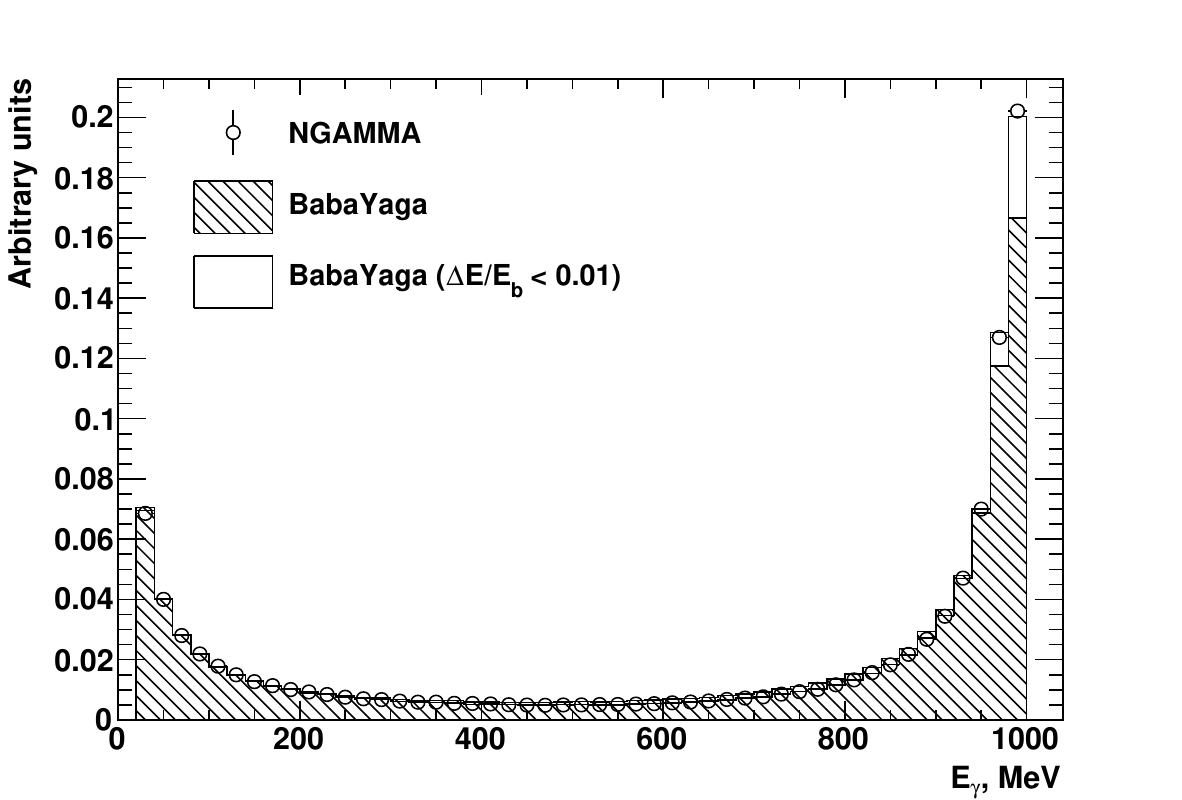}
\caption{\label{3gDist}
The distribution of photon energy $E_{\gamma}$ for the $e^+e^- \to 3\gamma$ events with $E= 2$ GeV.
The circles show the distribution obtained with the NGAMMA generator.
The solid histogram shows the distribution obtained with the BabaYaga@NLO generator with a constraint on 
the total energy of the additional photons $\Delta E < 0.01 \cdot E_{b}$.
The hatched histogram represents the same distribution without this constraint. All distributions are normalized to unity.
See text for a detailed description of the applied cuts.
}
\end{figure}

The distribution of the photon energy is shown in Fig.~\ref{3gDist} for $e^+e^- \to 3\gamma$
events at the c.m. energy $E = 2$ GeV. The distribution generated according to~(\ref{NgammaM}) with all three photons 
satisfying~(\ref{SNDcut}) is plotted by circles. The hatched histogram shows the distribution 
for $e^+e^- \to 2\gamma$ events generated by BabaYaga@NLO 
for which exactly three photons satisfy~(\ref{SNDcut}). The solid histogram shows the distribution for the same 
events, but  with an additional cut on the total energy of the other photons in the event $\Delta E < 0.01 \cdot E_{b}$.
Since all distributions are normalized to unity, we compare 
only their shapes. The distributions for the Born amplitude and for BabaYaga@NLO 
with restricted additional radiation are in good agreement.
However, removing the constraint on $\Delta E/E_{b}$ notably changes the distribution for BabaYaga@NLO events.

We expect that, for $N > 2$, the theoretical accuracy of the Born approximation~(\ref{NgammaM}) remains 
the same order of magnitude and can be roughly estimated as 10\% for relatively soft kinematic restrictions.

\section{Approximate expression for the amplitude}

One of the main drawbacks of expression~(\ref{NgammaM}) given above is the rapid growth with $N$ of the number of helicity states~(S)
and the number of photon permutations~(D). As a consequence, this leads to a rapid increase in the time 
required to calculate the cross section with increasing $N$. The use of the spiral technique helps to curb this growth, but even in this case 
the number of terms in~(\ref{NgammaCrs1}) grows as $2^N N!$. For this reason, an approximate expression for the matrix 
element squared can be very useful~\cite{Kleiss_ng}
\begin{equation}
\bar{M}^2_{ap} = e^{2N} s^{N-2} \left[ \sum_{i=1}^{N}x_i y_i (x_i^2+y_i^2)\right]\prod_{j=1}^{N}(x_i y_i)^{-1}, 
\label{NgammaMap}
\end{equation}
\begin{equation*}
x_i = p_{+}\cdot k_i, \quad y_i = p_{-}\cdot k_i.
\end{equation*}
For $N = 2$ and $N = 3$ it is the exact expression for the summed/averaged amplitude squared. 
For $N \ge 4$ it gives the correct result only in the asymptotic limit
where $N-3$ photons become soft. 

To evaluate the accuracy of the approximate 
expression~(\ref{NgammaMap}), we performed a Monte Carlo study.
We generated $10^7$ events at the c.m. energy $E = 2$ GeV, uniformly distributed in the five-photon phase space with two constraints: 
the minimum photon energy in an event must be more than $E_{min} = 20$ MeV and the minimal
angle between the photon momenta and the beam axis must be more than $\theta_{min} = 15^{\circ}$.
The value of the spin summed/averaged matrix element squared was calculated for each event according to the exact ($\bar{M}^2$) and
approximate ($\bar{M}^2_{ap}$) expressions. For convenience, the matrix elements were multiplied by the integral 
over phase space and divided by the flux factor
\begin{equation}
\sigma_{N\gamma} = \frac{1}{F}\cdot \textstyle\int d\Phi_{N}\displaystyle\cdot\bar{M}^2, \qquad \hat{\sigma}_{N\gamma} = \frac{1}{F}\cdot \textstyle\int d\Phi_{N}\displaystyle\cdot\bar{M}^2_{ap},
\label{DiffCrs}
\end{equation}
\begin{equation*}
\textstyle\int d\Phi_{N} = \displaystyle \int \prod_{i=1}^{N}\frac{d^4k_i}{(2\pi)^3}\theta(k_i^0)\delta(k_i^2)(2\pi)^4\delta^4(P-\sum_{i=1}^{N}k_i)
\end{equation*}
\begin{equation}
= (2\pi)^{4-3N}\left(\frac{\pi}{2}\right)^{N-1}\frac{E^{2N-4}}{\Gamma(N)\Gamma(N-1)}.
\label{PSint}
\end{equation}

\begin{figure}[]
\center
\includegraphics[width=1.1\columnwidth]{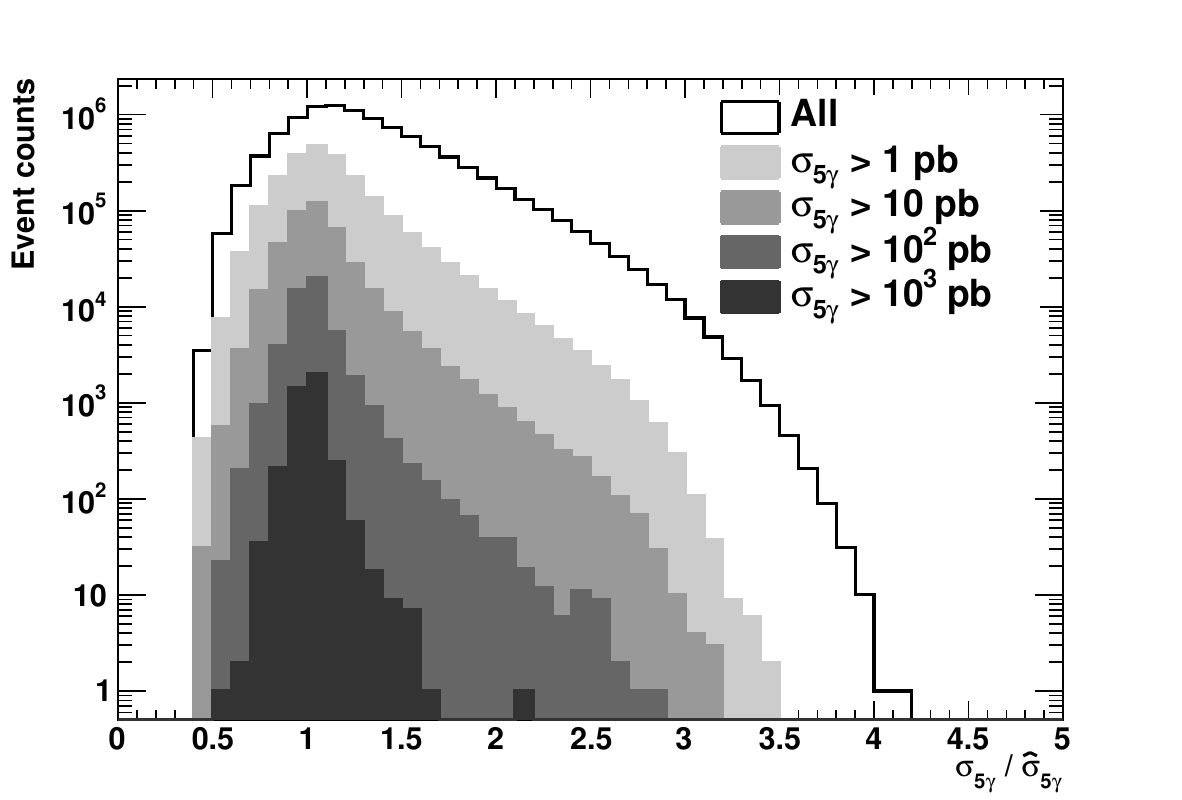}
\caption{\label{DiffCrsPic}
Distribution of $\sigma_{5\gamma} / \hat{\sigma}_{5\gamma}$ depending on the constraint on $\sigma_{5\gamma}$
for $e^+e^- \to 5\gamma$ events uniformly distributed in phase space with $E= 2$ GeV, $E_{min} = 20$ MeV and $\theta_{min} = 15^{\circ}$.
}
\end{figure}

Figure~\ref{DiffCrsPic} shows the distribution of the ratio
$\sigma_{5\gamma} / \hat{\sigma}_{5\gamma}$
for different constraints on $\sigma_{5\gamma}$. It can be seen that the accuracy of the approximate expression is not high.
At the same time, it increases significantly in the most important area where the cross section becomes larger $\sigma_{5\gamma} > 1$ nb. For this reason,
although the approximate expression does not provide a sufficient description for most of the phase space, it gives
reasonable accuracy for the total cross section and distributions of kinematic parameters.
But the most important thing is that, despite the variation of $\sigma_{5\gamma}$ by about ten orders of magnitude, $\hat{\sigma}_{5\gamma}$ differs 
from it only by a factor of several.

\section{Event generation strategy}

It can be seen from the approximate expression~(\ref{NgammaMap}) that 
the matrix element of the $e^+e^-\to N\gamma$ process has peaks near $x_i,y_i \approx 0$.
They correspond to the kinematics when some of the photons 
have energies and polar angles close to the minimum allowed values.
These peaks tremendously reduce the generation efficiency when using 
the standard acceptance-rejection method. 
Together with the time-consuming calculation of the exact matrix 
element~(\ref{NgammaM}), this gives a prohibitively long event generation time.
To overcome these obstacles, we split the generation procedure with the acceptance-rejection method into two stages.
At the first stage, we obtain a distribution according to $W(\Phi_{N})d\Phi_{N}$
from events uniformly distributed in the phase space~\cite{rambo}.
For the function $W$ we use $\hat{\sigma}_{N\gamma}$ truncated to some value $\sigma_{N\gamma}^{cut}$
\begin{equation*}
W = \hat{\sigma}_{N\gamma}~{\rm when}~\hat{\sigma}_{N\gamma} <  \sigma_{N\gamma}^{cut}~\rm{and}
\end{equation*}
\begin{equation}
W = \sigma_{N\gamma}^{cut}~{\rm when}~ \hat{\sigma}_{N\gamma} \ge  \sigma_{N\gamma}^{cut}.
\label{SigCut}
\end{equation}
At the second stage, we accept only those events which satisfy the condition
\begin{equation}
\sigma_{N\gamma}/W > \Delta\cdot r,
\label{SigCut2}
\end{equation}
where $r$ is a random number uniformly distributed in $(0,1)$, $\Delta$ is an adjustable parameter. 
It is convenient to choose $\sigma_{N\gamma}^{cut}$ and $\Delta$ so that $\sigma_{N\gamma}/W$ is less than $\Delta$
for a small fraction of the generated events. In this case, we need to supplement our procedure by assigning weights to the events.
For all events with $\sigma_{N\gamma}/W < \Delta$ we 
assign a unit weight $w = 1$, otherwise $w = \sigma_{N\gamma}/(\Delta \cdot W)$.

\begin{figure}[]
\center
\includegraphics[width=1.1\columnwidth]{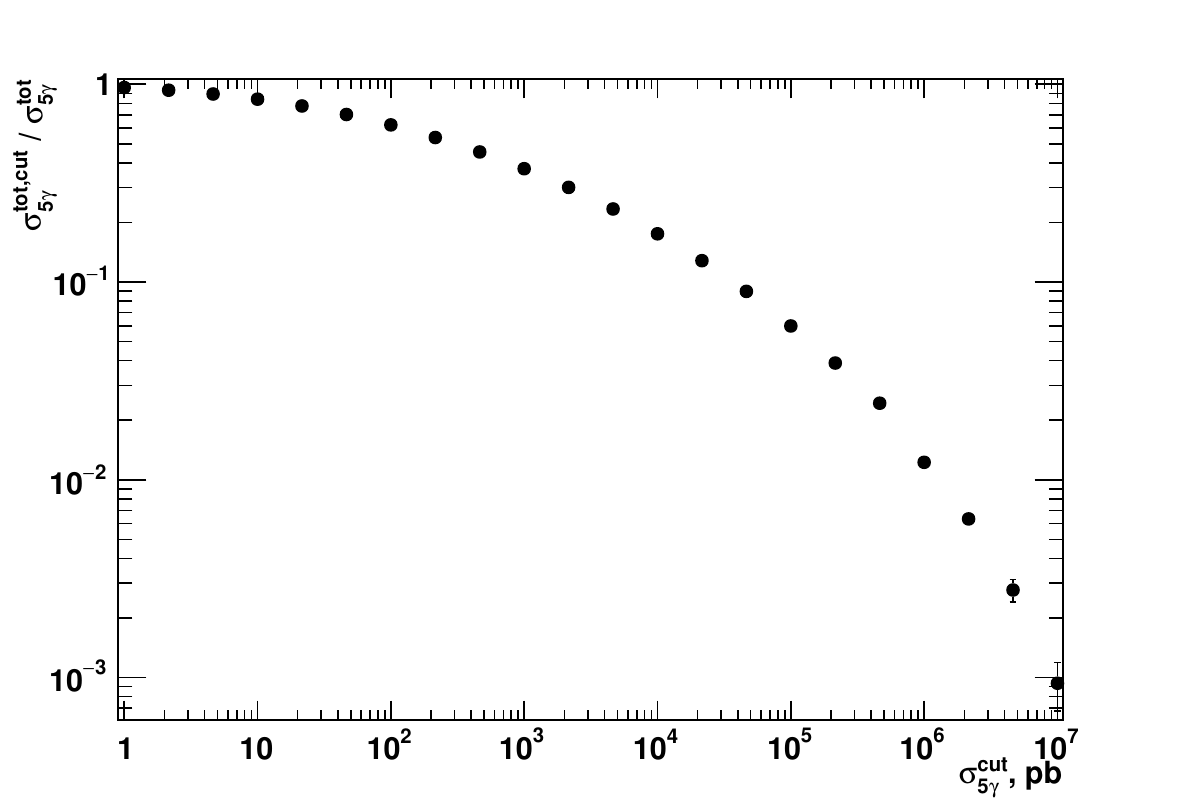}
\caption{\label{CrsCutPic}
Ratio of the total cross section integrated over the area where $\sigma_{5\gamma} > \sigma_{5\gamma}^{cut}$ ($\sigma_{5\gamma}^{tot, cut}$)
to the total cross section $\sigma_{5\gamma}^{tot}$ for $e^+e^- \to 5\gamma$ events
with $E= 2$ GeV, $E_{min} = 20$ MeV and $\theta_{min} = 15^{\circ}$.
}
\end{figure}

The first stage is the most inefficient and requires calculation of the matrix elements
for numerous points in the phase space. Using the approximate expression for it significantly reduces computation time. 
Taking a large enough $\sigma_{N\gamma}^{cut}$ we can minimize the fraction of events with non-unit weights to almost
negligible values and still obtain a gain in generation efficiency of several orders of magnitude.
Let us consider as an example the generation $e^+e^- \to 5\gamma$ events 
with $E= 2$ GeV, $E_{min} = 20$ MeV and $\theta_{min} = 15^{\circ}$.
From Fig.~\ref{DiffCrsPic} one can see that the inequality $\sigma_{5\gamma}/\hat{\sigma}_{5\gamma} < \Delta$ is valid
for the entire phase space if we take $\Delta = 5$. 
In this case, the phase space corresponding to
events with $w > 1$ is the area where $\sigma_{5\gamma} > \Delta \cdot \sigma_{5\gamma}^{cut}$.
From Fig.~\ref{CrsCutPic} one can see that if we take $\sigma_{5\gamma}^{cut} = 1$ $\rm\mu$b, 
the fraction of the total cross section corresponding to
this area is $\approx 0.3 \%$. At the same time, maximum of $\sigma_{5\gamma}$ in the allowed phase space is $\sigma_{5\gamma}^{max} = 4.8$ mb,
and if we take $\sigma_{5\gamma}^{cut} =\sigma_{5\gamma}^{max}$, then the generation efficiency will drop by more than three orders of magnitude.

\section{Comparison with CompHEP}

\begin{figure*}[]
\center
\includegraphics[width=0.45\textwidth]{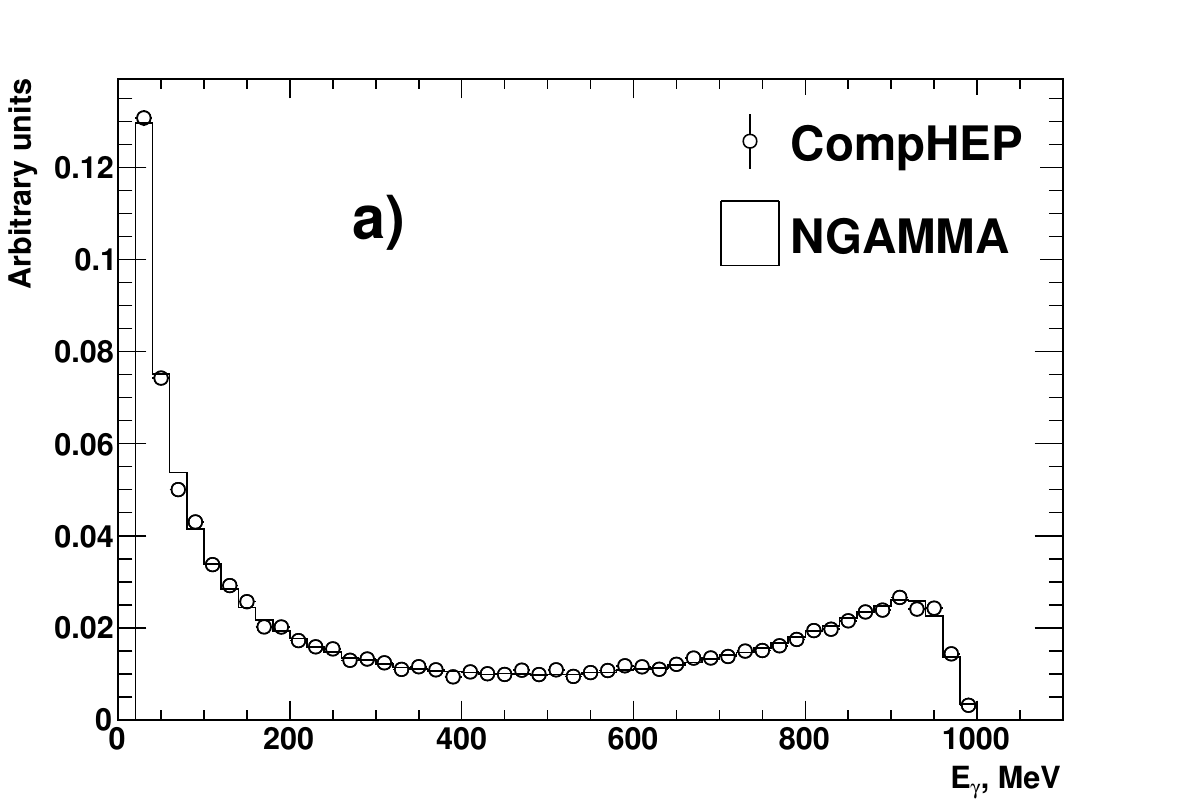}\hfill 
\includegraphics[width=0.45\textwidth]{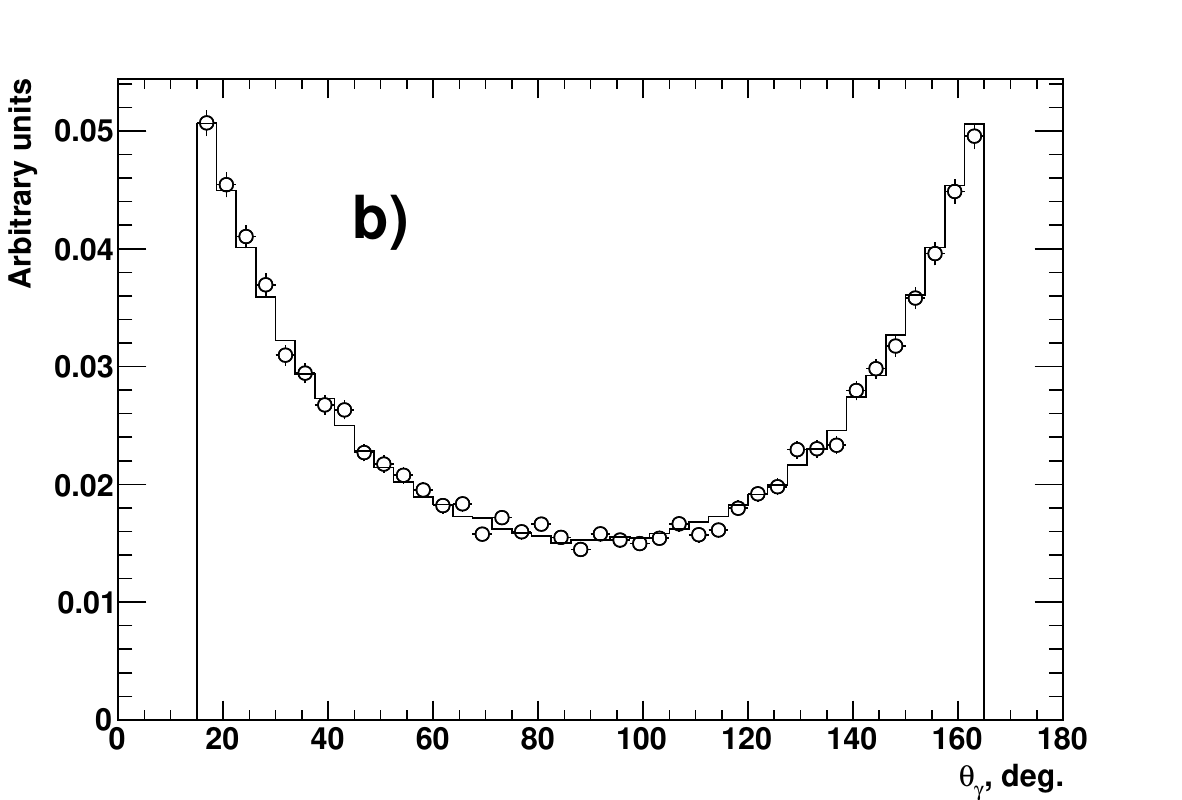}\\
\includegraphics[width=0.45\textwidth]{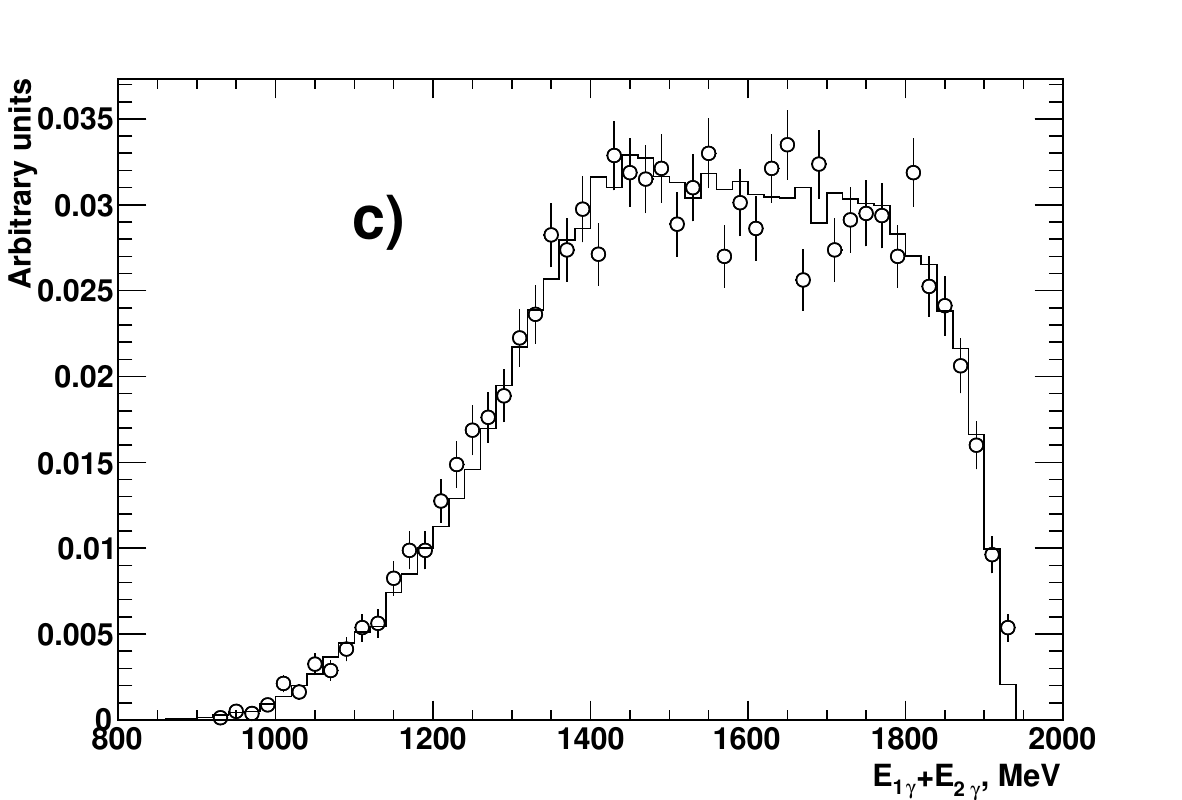}\hfill
\includegraphics[width=0.45\textwidth]{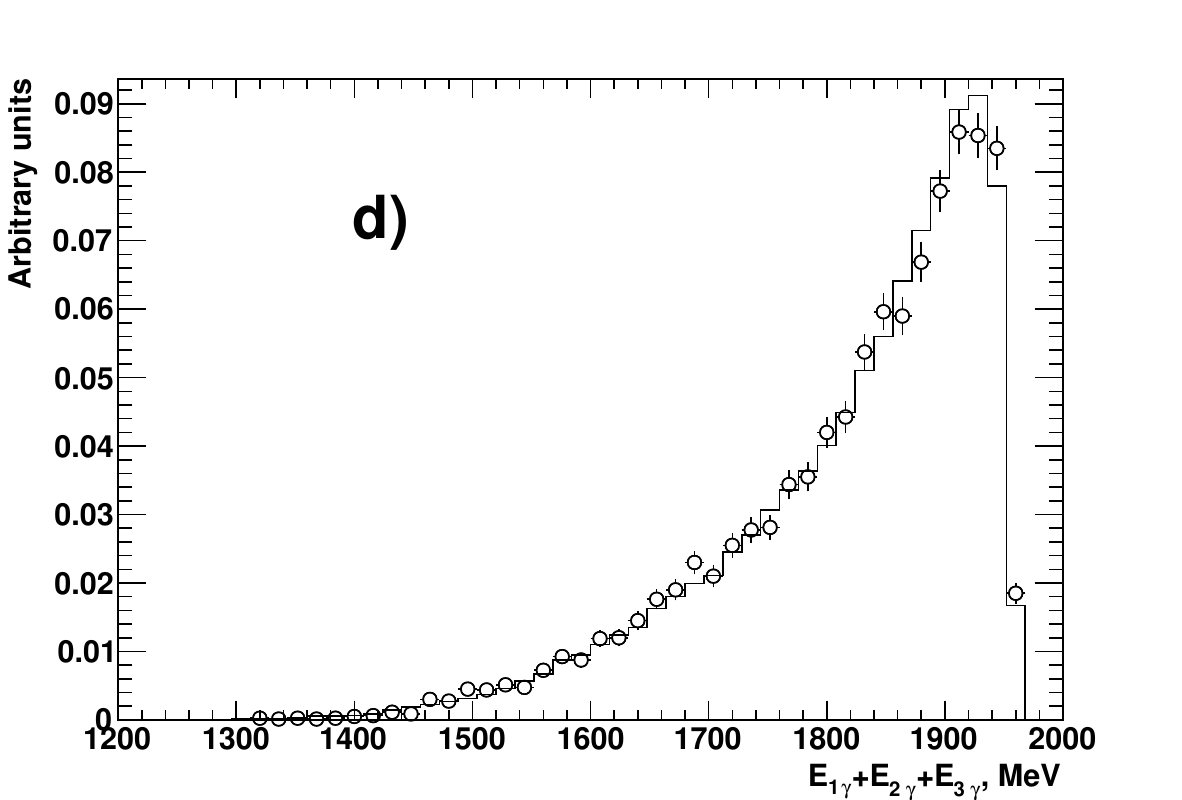}
\caption{\label{5gDist}
Comparison of distributions for the $e^+e^- \to 5\gamma$ events with $E= 2$ GeV,
$E_{min} = 20$ MeV, $\theta_{min} = 15^{\circ}$ obtained with NGAMMA (solid histogram) and CompHEP (circles).
a) Photon energy $E_{\gamma}$;
b) Photon polar angle $\theta_{\gamma}$;
c) Sum of energies of the two most energetic photons $E_{1\gamma}+E_{2\gamma}$;
d) Sum of energies of the three most energetic photons $E_{1\gamma}+E_{2\gamma}+E_{3\gamma}$
}
\end{figure*}

To examine the accuracy and performance of the NGAMMA generator, we perform a comparison with the CompHEP package~\cite{CompHEP}.
The distributions of the kinematic parameters for both programs for $e^+e^- \to 5\gamma$ with $E_{min} = 20$ MeV, 
$\theta_{min} = 15^{\circ}$ are shown in Fig.~\ref{5gDist}. As one can see, they are in good agreement with each other everywhere
except at the very right end of the distribution in Fig.~\ref{5gDist}c. 
This region corresponds to the kinematics in which two photons 
accumulate almost all the energy of the event. The cross section in this region has a singular behavior 
which, as discussed above, creates significant complications for event generation.

\begin{figure*}[]
\center
\includegraphics[width=0.45\textwidth]{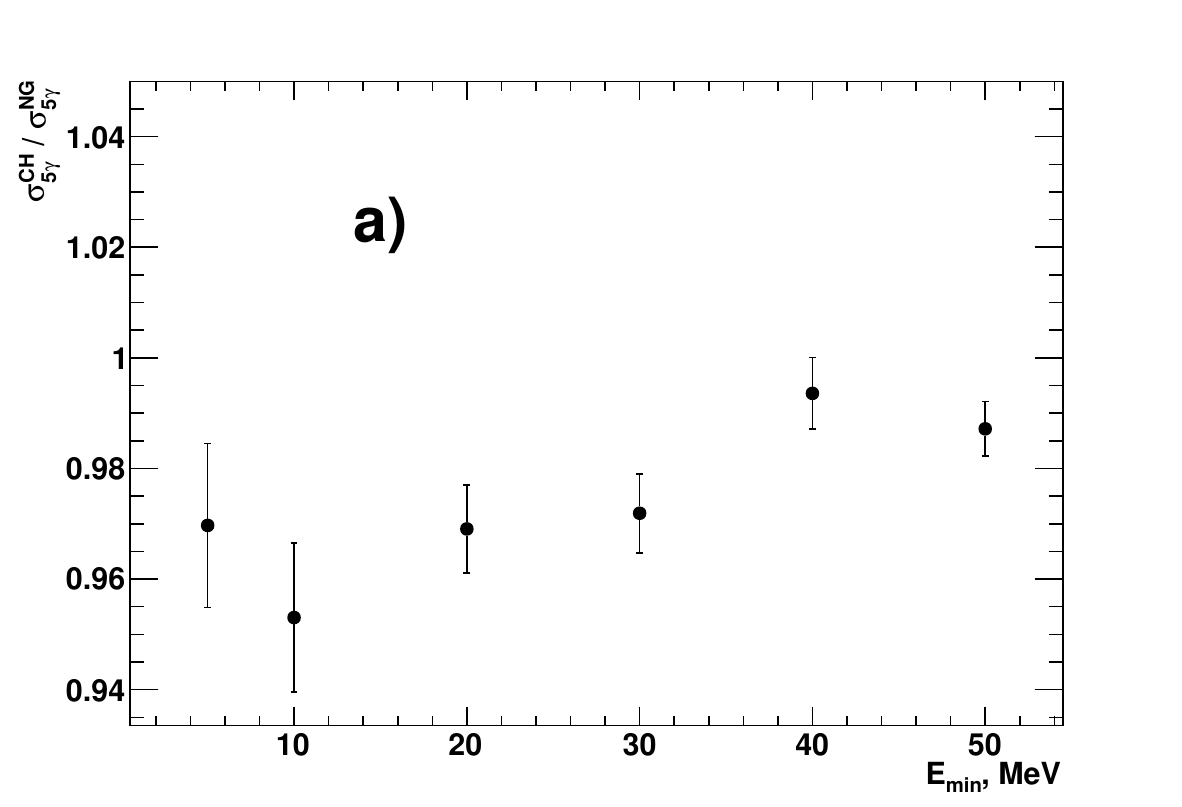}\hfill 
\includegraphics[width=0.45\textwidth]{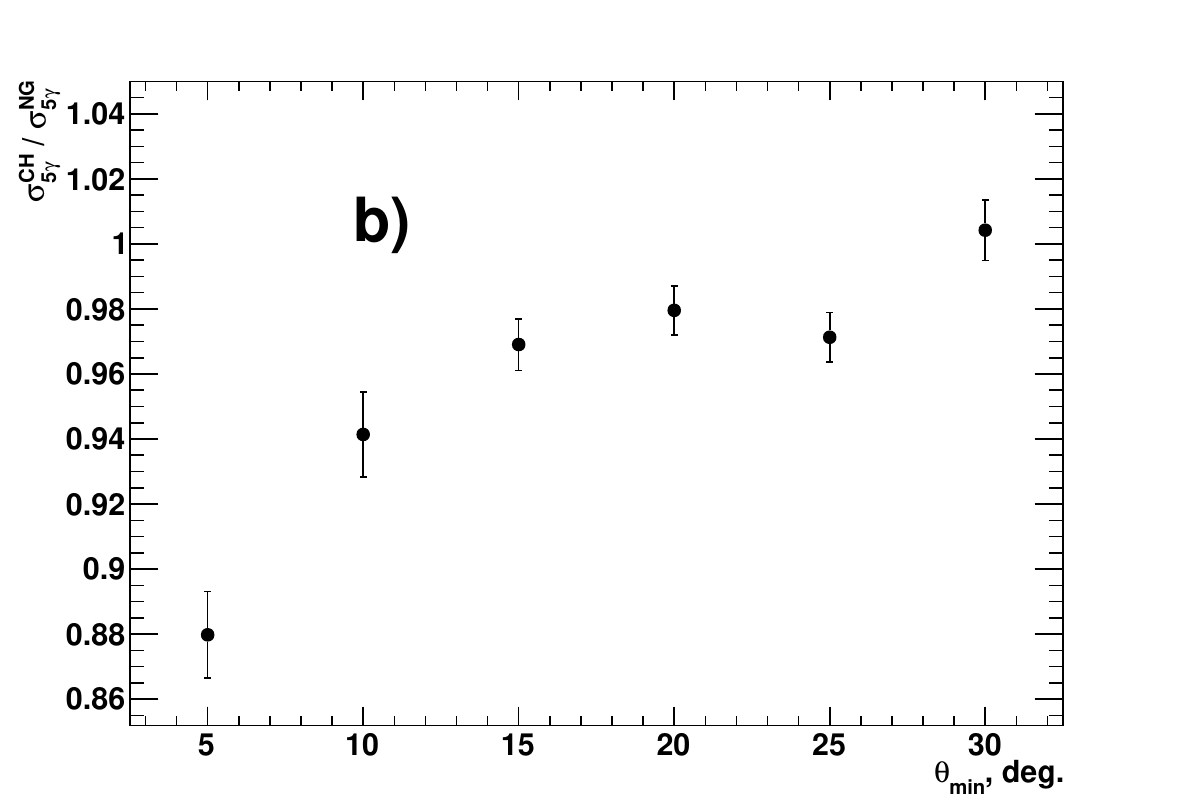}\\
\caption{\label{CrsET}
The ratio of $e^+e^- \to 5\gamma$ cross sections calculated using 
NGAMMA ($\sigma_{5\gamma}^{NG}$) and CompHEP ($\sigma_{5\gamma}^{CH}$).
a) Dependence of the ratio on  $E_{min}$ with $\theta_{min} = 15^{\circ}$;
b) Dependence of the ratio on $\theta_{min}$ with $E_{min} = 20$ MeV.
}
\end{figure*}

To check the accuracy of the total cross section calculation, we calculate it for $e^+e^- \to 5\gamma$ with different 
constraints on the energies and angles of the photons. The ratio of calculated cross sections with CompHEP and NGAMMA
is shown in Fig~\ref{CrsET}. A clear dependence of this ratio on $\theta_{min}$ is visible.
The total cross sections coincide at $\theta_{min} = 30^{\circ}$.
The difference between them grows with decreasing $\theta_{min}$ and reaches 12~\% at $\theta_{min} = 5^{\circ}$.

For comparison of event generation performance, we generate events of $e^+e^- \to 5\gamma$ with 
$E_{min} = 20$ MeV, $\theta_{min} = 15^{\circ}$ using
a computer running on an Intel Xeon E5-2620 series CPU operating at 2 GHz. 
CompHEP requires a preparatory stage including cross section calculation and search for the cross section maximum.
With default settings this preparatory stage takes about 12 hours and generation of 100 events takes another 8 hours.
For NGAMMA with $\Delta = 5$ and $\sigma_{5\gamma}^{cut} = 1$ $\rm\mu$b generation of 100 events takes 10 minutes.

\section{Description of the NGAMMA software}
NGAMMA is a C++ software with a simple code structure of two main components: the \texttt{NGammaGenerator} class and the \texttt{startgen.cc} wrapper. 
\texttt{NGammaGenerator} holds all generation logic described in the previous sections, while the
wrapper responsibility is to configure the \texttt{NGammaGenerator} instance according to the user 
input settings, run the event loop and save the output data. 

The following methods of the \texttt{NGammaGenerator} class are the most important for the generator:
\noindent
\begin{description}[style=nextline] 
	\item[doEvent]  is the main method responsible for one event generation,
	\item[checkCut] checks if the current event is inside the  allowed phase space defined by the following conditions on each photon in the event
	\begin{equation}
		E_{\gamma} > E_{min}, \qquad    \theta_{min} < \theta_{\gamma} < 180^{\circ}-\theta_{min}.
		\label{NGcut}
	\end{equation}
	\item[findMajorant] calculates the default value of the majorant at the
	first stage of generation as $\hat{\sigma}_{N\gamma}$ corresponding to the kinematics
	in which two photons hold most of the event energy and other photons have energies close to the minimal allowed  value $E_{min}$.
\end{description}

If needed, one can use C++ inheritance mechanism to modify and extend \texttt{NGammaGenerator} functionality with minimal changes to the wrapper. 

\subsection{Installation instructions and the test run}
NGAMMA is supported on Linux. It has been successfully compiled and tested on Alma Linux 9~\cite{almalinux}.  

Building NGAMMA from source with the CMake tool~\cite{cmake} is the only available installation method. 

The following dependencies are required to be installed:
\begin{itemize}
	\item CMake tool~\cite{cmake} version 3.26 or newer;
	\item CERN ROOT framework~\cite{root} version 6.32 or newer;
	\item Boost libraries~\cite{boost} version 1.75 or newer.
\end{itemize}

It is worth noting that the listed above minimal versions of the dependencies set up the environment known to be suitable for NGAMMA. Other options have not been tested. 

The NGAMMA itself is written in C++11 but it depends on the ROOT framework and must be compiled with the same standard which was used for  ROOT compilation 
(it is C++17 for version 6.32). So to choose an appropriate compiler one must take into account a C++ standard officially supported by the chosen ROOT version.  

To build and install NGAMMA one needs to run the following commands:
\begin{lstlisting}[language=bash,breaklines=true]
$ mkdir <buildDir> <installDir>
$ cd <buildDir>
$ cmake  -DCMAKE_INSTALL_PREFIX=~\linebreak~<installDir> <sourceDir>
$ cmake --build . --target install
\end{lstlisting}

Here \texttt{buildDir} and \texttt{installDir} are directories for the build and for the installation correspondingly, 
while \texttt{sourceDir} is a directory containing the NGAMMA sources. The above commands install the NGAMMA 
executable \texttt{ngammagen} in the directory \texttt{installDir/bin} and the NGAMMA default configuration file \texttt{genconfig.cfg} in \texttt{installDir}. 

To check the installed program, one can run the NGAMMA executable with the settings from the installed configuration file:
\begin{lstlisting}[breaklines=true]
$ cd <install_dir> 	
$ ./bin/ngammagen 
\end{lstlisting}

This should produce an output similar to the content of the file \texttt{test.log} with input generator parameter values and NGAMMA final statistics 
and the text file called \texttt{test.txt} with generated events. Both files are provided with the NGAMMA sources.

\subsection{NGAMMA settings}
There are two sets of parameters available for a user to set up. The first one is generator 
parameters used for the \texttt{NGammaGenerator} instance configuration. The second set is 
used for the wrapper. All parameters are listed in a help message when NGAMMA is called with `\texttt{--help}' option.

The generator parameters are described below:
\begin{description}[style=nextline] 
	\item[rndseed] is a seed for the random number generator;
	
	\item[nevents] is the number of generated events;
	
	\item[ecm] is the center of mass energy of the initial $e^{+}e^{-}$ $E$ (GeV);
	
	\item[ng] is the number of photons in the final state $N$;
	
	\item[emin] is the minimal photon energy $E_{min}$ (GeV);
	
	\item[tmin] is the minimal photon polar angle $\theta_{min}$ (deg.);
	
	\item[maj1] is the parameter $\sigma_{N\gamma}^{cut}$ in equation (\ref{SigCut});
	
	\item[maj2] is the parameter $\Delta$ in equation (\ref{SigCut2}).
	
\end{description}

The wrapper parameters control input and output settings:
\begin{description}[style=nextline] 
	\item[verbose] is a logging level;

	\item[config] is the name of the configuration file;
	
	\item[ofileFormat] is the output file format to store photons (root or txt);
	
	\item[ofileName] is  the output file name to store photons (without extension).
\end{description}

All generator parameters can be set using both the configuration file and command line options. 
Most of the wrapper parameters can be set only with command line options, but \texttt{ofileFormat} 
and \texttt{ofileName} can be also set in the configuration file.
All configuration file parameters can be overridden with command line options (`\texttt{--<parameter name>}'). 
If some parameter is set neither in the configuration file nor with command line options, then the hardcoded value is used. 
These values can be found in the help message.	

\subsection{Input and output files}
An input configuration file is required for running the wrapper. This file contains the values of the generation parameters in the key-value \texttt{INI} format.
By default, the configuration file is named `\texttt{genconfig.cfg}' and expected to be in the current working directory. 
The configuration file can be specified  with the `\texttt{-c}' option.

NGAMMA saves all generated events into an output file. 
The wrapper supports two output file formats: a \texttt{text} file and a \texttt{ROOT} file with \texttt{ROOT::TTree} inside.
The format can be specified with the `\texttt{-ofileFormat}' option. The extensionless  output file name can be set using 
the `\texttt{-ofileName}' option. By default, the output file is created as a text file with an auto-generated name based 
on the main  generator parameters and the current date.

In case of the text file output each event is stored as one line with space separated fields 
in the following order: event number, event weight, 
$k_{1}^{x}$, $k_{1}^{y}$, $k_{1}^{z}$, $k_{1}^{0}$, \ldots , $k_{N}^{x}$, $k_{N}^{y}$, $k_{N}^{z}$, $k_{N}^{0}$.

If the \texttt{ROOT} format is selected for the output file then the \texttt{TTree} object is created with the following branches:
\begin{description}
	\item[w]  is the event weight;
	\item[ng]  is the number of photons in the final state;
	\item[kx]  is the array of 4-vector x components $k^{x}$ for all photons in the event;
	\item[ky]  is the same as \texttt{kx} but for y component $k^{y}$;
	\item[kz]  is the same as \texttt{kx} but for z component $k^{z}$;
	\item[k0]  is the same as \texttt{kx} but for time component $k^{0}$.
\end{description}

These branches are filled per each output event and the \texttt{TTree} object is written to the ROOT file after the event loop is finished.

\section{Summary}
The Monte Carlo event generator NGAMMA, for simulation of QED processes $e^+e^-\to N\gamma~(N \ge 2)$, has been developed.
It allows to generate momenta of outgoing particles and calculate the total cross section of the reaction.
For the generation it exploits a tree-level amplitude calculated using the spinor technique. The NGAMMA generator shows much better 
performance in comparison with the CompHEP package.

\section{CRediT authorship contribution statement}
L.~V.~Kardapoltsev: Writing – review \& editing, Writing – original draft, Supervision, Software, Methodology,
Formal analysis, Conceptualization. N.~A.~Melnikova: Writing – review \& editing, Writing – original
draft, Software, Validation, Methodology. 

\section{Declaration of Competing Interest}
The authors declare that they have no conflict of interest regarding this manuscript.

\section{Acknowledgments}

The authors are grateful to A.A.~Korol, R.N.~Lee and A.E.~Obrazovsky for useful discussions.

\end{document}